\documentstyle[amssymb,multicol,amstex,aps,,epsfig]{revtex}

\begin{document}
\title{Charge Order in NaV$_2$O$_5$ studied by EPR}
\author{M.~Lohmann$^1$, H.-A.~Krug von Nidda$^1$, M. V. Eremin$^{1,2}$, A.~Loidl$^1$}
\address{$^1$Experimentalphysik V, EKM, Institut f\"{u}r Physik, Universit\"{a}t Augsburg,\\ D-86135 Augsburg,
Germany \\
 $^2$Kazan State University, 420008 Kazan, Russia}
\author{G. Obermeier, S. Horn}
\address{Experimentalphysik II, Institut f\"{u}r Physik, Universit\"{a}t Augsburg,\\ D-86135 Augsburg,
Germany}
\date{\today}
\maketitle

\begin{abstract}
We present angular dependent EPR measurements in NaV$_2$O$_5$ at X-band frequencies in the
temperature range $4.2\,$K $\leq T \leq 670$\,K. A detailed analysis in terms of the antisymmetric
Dzyaloshinski-Moriya and the anisotropic exchange interactions yields the following scheme of
charge order: On decreasing temperature a quarter-filled ladder with strong charge disproportions,
existing for $T\geq 100$\,K, is followed by zig-zag charge-order fluctuations which become
long-range and static below $T_{\rm{SP}}=34$\,K.\\
\end{abstract}

\pacs{PACS numbers:  75.10.Jm, 75.30.K, 75.45.+j, 76.30.-v}

\begin{multicols}{2}

The observation of an exponential decrease in magnetic susceptibility in $\alpha$'-NaV$_2$O$_5$
below 34\,K \cite{Iso:96} triggered many experimental and theoretical investigations. Following the
determination of the crystal structure by Carpy and Galy \cite{Carpy:75} as space group P$2_1mmn$
with linear chains of V$^{4+}$ ions (spin $S=1/2$) separated by non magnetic V$^{5+}$ chains,
NaV$_2$O$_5$, analogous to CuGeO$_3$, was classified as inorganic spin-Peierls system. Later on a
re-investigation of the crystal structure showed that NaV$_2$O$_5$ has to be considered a
quarter-filled ladder system \cite{Smo:98,vonSchn:98,Mee:98} with only one vanadium site in the
high-temperature phase described by space group P$mmn$. The direction of the ladders is given by
the crystallographic $b$ axis. Each rung consists of two VO$_5$ pyramids along the $a$ axis, which
share one corner at their base. Based on this structure a charge-order transition followed by
different kinds of spin order were proposed \cite{Smo:98,Thal:98,Seo:98}. A recent investigation
suggested a low-temperature phase consisting of modulated ladders with zig-zag charge order
alternating with non modulated ladders with vanadium ions of intermediate valence V$^{4.5+}$
\cite{Lued:99}.\\

Electron-paramagnetic resonance (EPR) is ideally suited for the investigation of the magnetic
properties of vanadium systems, since the vanadium ions themselves (as V$^{4+}$ with electron
configuration $3d^1$ in the present case) can be used as a microscopic probe of the spin system. In
this paper we present angular dependent EPR measurements at X-band frequencies (9.48\,GHz) within
the temperature range 4.2\,K $<T<670$\,K. The experimental details concerning the crystal growth of
NaV$_2$O$_5$ and EPR measurements have been described in previous papers \cite{Loh:97}. For all
temperatures the EPR spectrum consists of a single strongly exchange-narrowed Lorentzian line. The
EPR intensity does not show any angular dependence within experimental error. Its temperature
behavior follows that of a one dimensional antiferromagnetic Heisenberg chain at high temperatures
$T> 100$\,K \cite{Bonner:64}, whereas it decreases exponentially below the transition temperature
$T_{\rm{SP}}$. Strong deviations from the model predictions were observed between 100\,K and
$T_{\rm{SP}}$ \cite{Loh:97}. The $g$-values are found as $g_{\rm{a}}\approx g_{\rm{b}}\approx 1.98$
and $g_{\rm{c}}\approx 1.94$, which are typical for V$^{4+}$ in octahedral crystal symmetry
\cite{Oga:86}, and increase only slightly below $T_{\rm{SP}}$. Here we confine ourselves to the
discussion of the angular and temperature dependence of the resonance linewidth, which carries the
information about the interactions of the vanadium spins with their local environment.

\begin{figure}
\begin{center}
\centerline{\epsfig{file=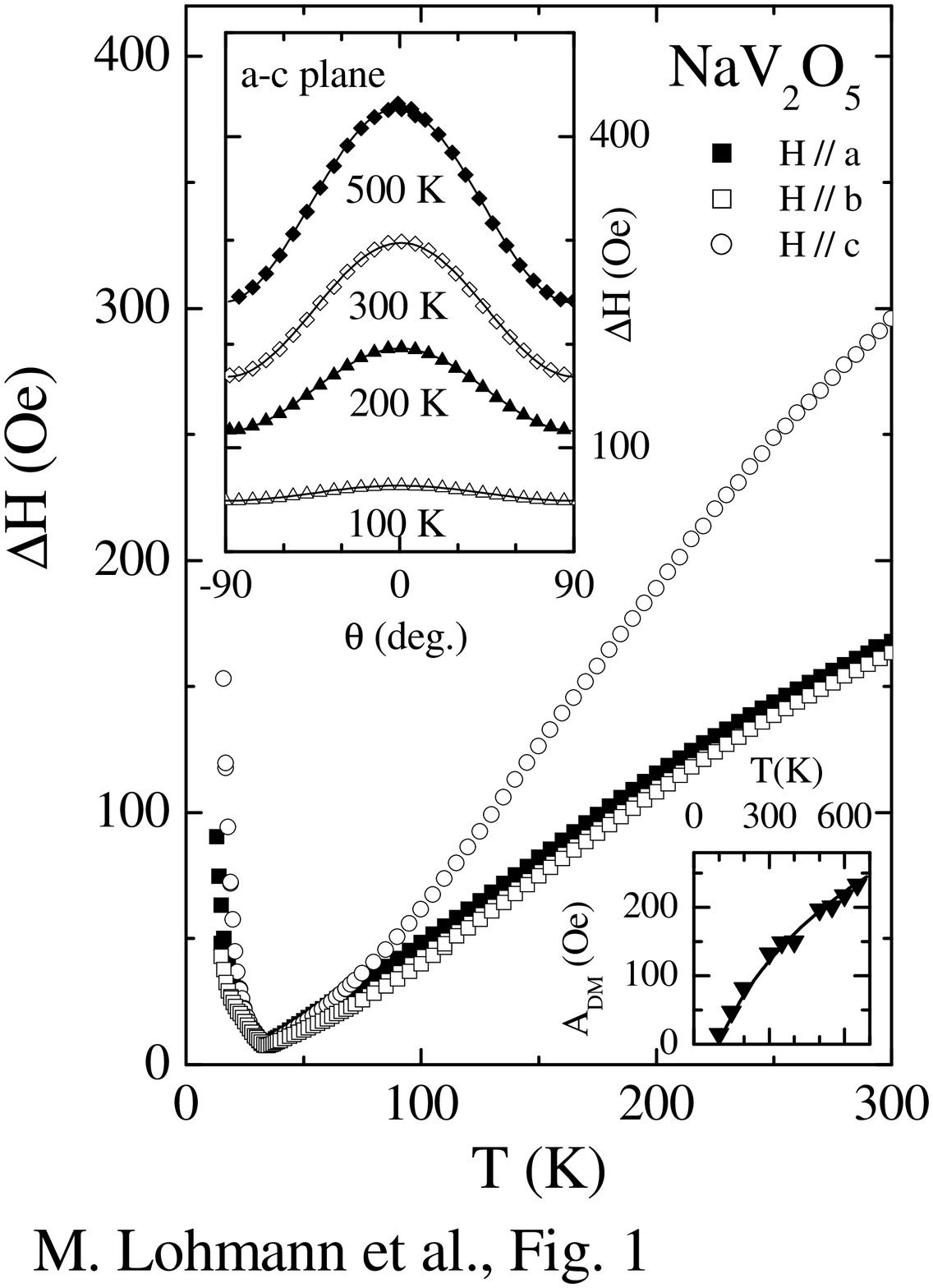,angle=0,clip,width=7cm}} \vspace{10pt}
\caption{EPR linewidth $\Delta H$ as function of temperature $T$
for the magnetic field applied along the three crystallographic axes. Upper inset: Angular
dependence of the EPR linewidth $\Delta H$ for the magnetic field applied perpendicular to the $b$
axis at temperatures $T\geq100$\,K. Lower inset: Temperature dependence of the strength parameter
$A_{\rm{DM}}$ of the DM interaction (see equation \ref{theta} and text).}
\label{fig1}
\end{center}
\end{figure}

Figure \ref{fig1} shows the temperature dependence of the linewidth $\Delta H$ with the external
magnetic field applied along the three main axis of the crystal. Starting with an isotropic value
$\Delta H = 8$\,Oe at the transition temperature $T_{\rm{SP}} = 34$\,K, the linewidth increases
monotonously with increasing temperature and develops a remarkable anisotropy with respect to the
$c$ axis of the crystal. In the respective temperature regime the curvature of $\Delta H(T)$ starts
with a positive sign and changes to negative values above 100\,K. Towards low temperatures
$T<T_{\rm{SP}}$ the line broadens anisotropically again probably due to the unresolved hyperfine
structure or due to a spin-glass transition \cite{Fud:99}.

It has been reported by Yamada {\em et al.} \cite{Yam:98} that the line broadening in the
high-temperature regime $T>100$\,K must be due to the antisymmetric Dzyaloshinski-Moriya (DM)
interaction, which is the only possible mechanism to explain the observed order of magnitude of
some 100\,Oe for the linewidth and its anisotropy. Estimations for anisotropic exchange and
dipole-dipole interactions yield values which are 100 and 1000 times smaller, respectively. The
angular dependence of the linewidth is shown in the upper inset of figure \ref{fig1} at four
different temperatures, where the $b$ axis was taken as rotation axis perpendicular to the static
magnetic field.
Following Yamada the data are well described by
\begin{equation}
\Delta H = A_{\rm{DM}} \cdot (1+\cos^2 \vartheta) + \Delta H_{0}
\label{theta}
\end{equation}
where $\vartheta$ is the polar angle with respect to the $c$ axis. $A_{\rm DM}$ is proportional to
the strength of the DM interaction and $\Delta H_{0}$ the residual linewidth due to further
relaxation mechanisms. Above 100\,K the parameter $A_{\rm{DM}}$ strongly increases, as it is shown
in the lower inset of figure \ref{fig1} and the experimental ratios $\Delta H_{\rm{c}}/\Delta
H_{\rm{a}}$ and $\Delta H_{\rm{c}}/\Delta H_{\rm{b}}$ nearly approximate the value of 2,
underlining the dominant influence of the DM interaction.

\begin{figure}
\begin{center}
\centerline{\epsfig{file=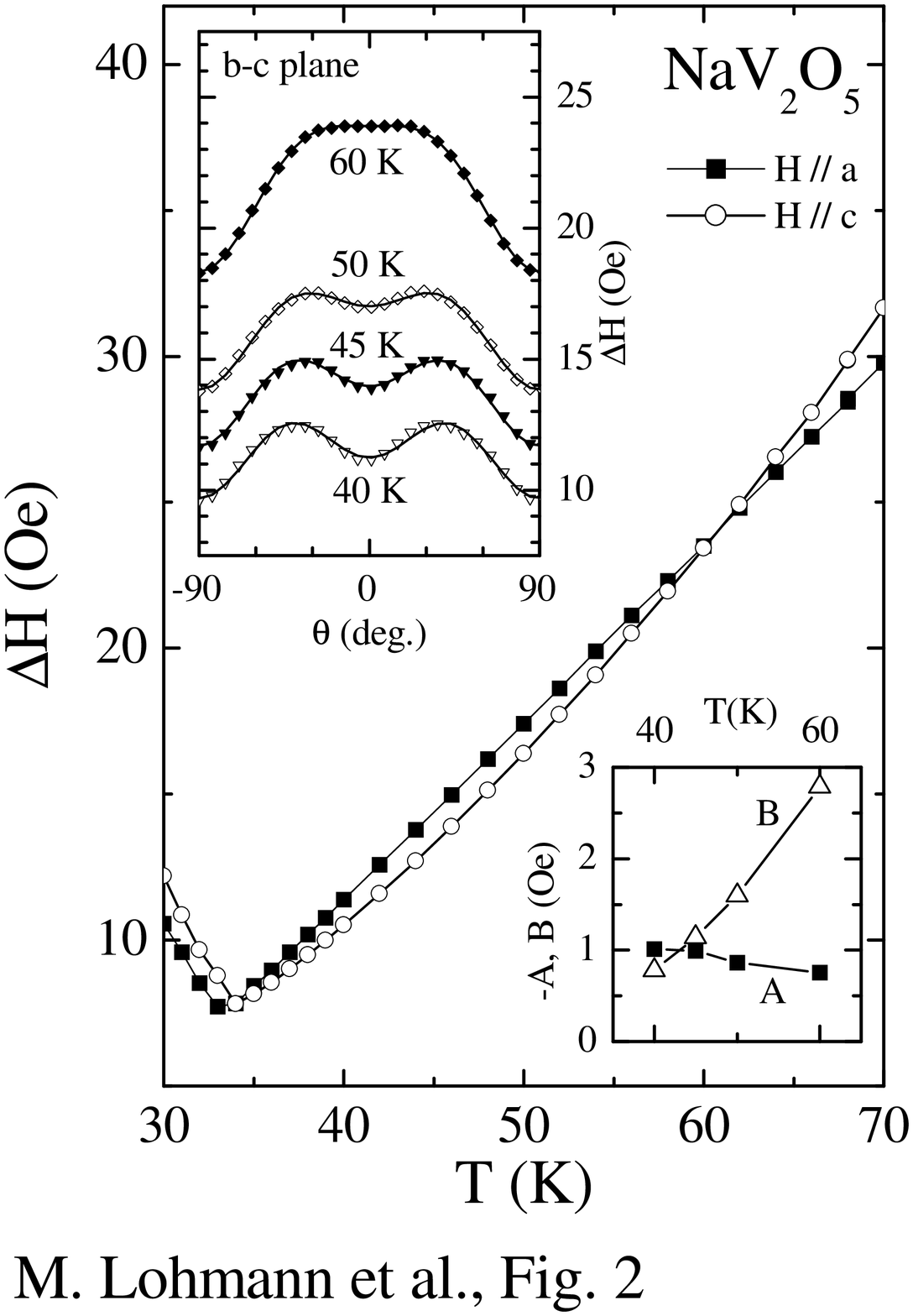,angle=0,clip,width=7cm}} \vspace{10pt}
\caption{Cross-over regime in the temperature dependence of the EPR
linewidth $\Delta H$. Upper inset: Angular dependence of the linewidth $\Delta H$ for the magnetic
field applied perpendicular to the $a$ axis at temperatures $T\leq100$\,K. Lower inset: Fit
parameters $A$ and $B$ derived from equation \ref{aniso} as function of temperature $T$.}
\label{fig2}
\end{center}
\end{figure}

Below $T=100$\,K the results strongly deviate from the high-temperature behavior. The positive
curvature of the linewidth indicates a change of the dominant interaction. In the temperature range
34\,K $<T<60$\,K, the maximum linewidth is not found for the magnetic field applied in $c$
direction any more, but it appears with respect to the $a$ axis. Figure \ref{fig2} nicely shows
this cross-over regime. The angular dependence within the $b$-$c$ plane shown in the upper inset of
figure \ref{fig2} is of special interest, because here an additional modulation appears, which is
well described by
\begin{equation}
\Delta H = A \cos(4 \vartheta) + B \cos(2 \vartheta) + \Delta H_{0}
\label{aniso}
\end{equation}
The amplitude $A$ of the modulation is about 1\,Oe, which is of the order of magnitude estimated
for the anisotropic exchange interaction, and increases slightly with decreasing temperature,
whereas the prefactor $B$, which can be considered to be due to the DM interaction, using
$\cos(2\vartheta)= 2\cos^2\vartheta-1$, as discussed below, strongly decreases.\\

The angular dependence of the EPR linewidth contains important information about the local
electronic distribution on the vanadium ladders. We start our discussion with respect to the
high-temperature regime $T>100$\,K, where the linewidth is determined by the DM interaction
\begin{equation}
{\cal H}_{\rm{ij}} = \bold{d_{ij}}\cdot[\bold{S_i} \times
\bold{S_j}]. \label{DMham}
\end{equation}
The vanadium spins $\bold{S_i}$ and $\bold{S_j}$ are coupled by superexchange via an oxygen ion.
The direction of the DM vector $\bold{d_{ij}}$ is determined by
\begin{equation}
{\bold d_{ij}} = d_{0}\cdot[{\bold n_{iO}} \times {\bold n_{Oj}}],
\label{DMv}
\end{equation}
where the space vectors $\bold{n_{iO}}$ and $\bold{n_{Oj}}$ connect the spins i and j with the
oxygen-bridge ion respectively. The maxi\-mum linewidth, which assigns the direction of the DM
vector, is found for the magnetic field applied parallel to the crystallographic $c$ axis.
\begin{figure}
\begin{center}
\centerline{\epsfig{file=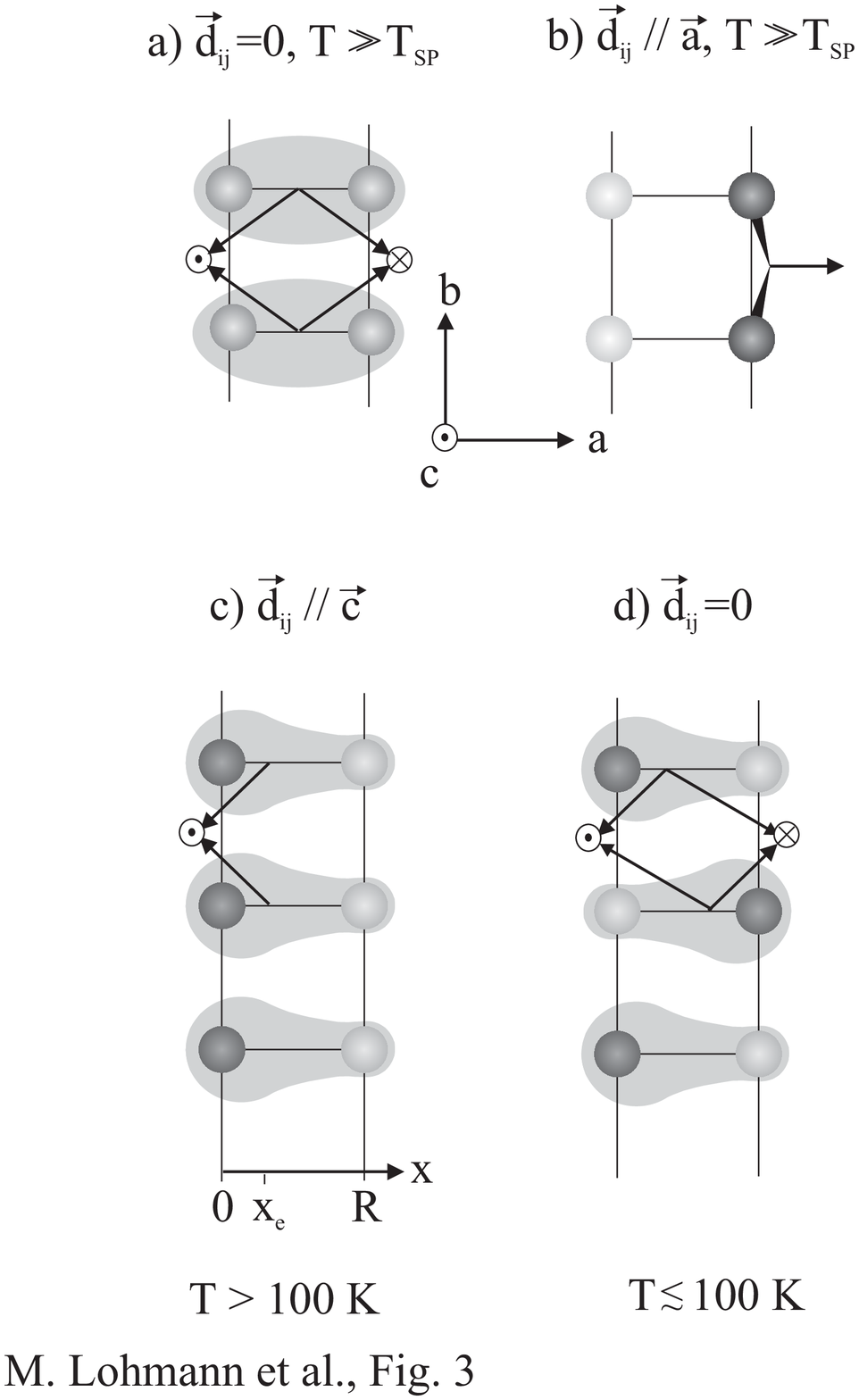,angle=0,clip,width=7cm}} \vspace{10pt}
\caption{Projection of the ladder structure of VO$_5$ pyramids into
the $a$-$b$ plane for different possible charge distributions, the spheres denote the vanadium
ions, the oxygen ions have been omitted for simplicity. The oxygen ions of the relevant V-O-V
bridges are marked by the DM vectors ${\bold d_{ij}}$. (a) Symmetric charge distribution (grey
clouds) of a quarter-filled spin ladder V$^{4.5+}$; (b) chains of V$^{4+}$ (dark) and V$^{5+}$
(light); (c) local charge disproportion at high temperatures; (d) charge order fluctuations
(dynamic, short-range) above $T_{\rm{SP}}$ and charge order (static, long-range) below
$T_{\rm{SP}}$. Case (c) and (d) are in full agreement with the EPR results.}
\label{fig3}
\end{center}
\end{figure}
Therefore the relevant oxygen-bridge vectors $\bold{n_{iO}}$ and $\bold{n_{Oj}}$ must build an
angle smaller than $180^{\circ}$ within the $a$-$b$ plane. Moreover the existence of a non
vanishing DM interaction requires a unit cell without inversion center, because otherwise the DM
vectors cancel each other, as it is the case for a quarter-filled spin ladder, where each vanadium
spin is equally distributed between two V$^{5+}$ ions on the rungs of the ladder (cf. Fig.
\ref{fig3}a). The structure determined by Carpy and Galy \cite{Carpy:75} fulfills the condition of
asymmetry, because of the separation of the vanadium ladders in V$^{4+}$ and V$^{5+}$ chains. Here
the DM interaction takes place only via one oxygen bridge between two neighboring V$^{4+}$ sites
along the $b$ axis (cf. Fig. \ref{fig3}b). However, using the structural data \cite{vonSchn:98} for
the vanadium and oxygen ions under consideration, we find that, with respect to the vanadium
positions, the oxygen ions within the chains are farther displaced along the $c$ axis $\Delta
c\approx 0.58\,{\rm\AA}$ than along the $a$ axis $\Delta a\approx 0.28\,{\rm\AA}$. This yields a
dominant contribution of the DM vector pointing into the $a$ direction, which is not observed
experimentally. We conclude that the EPR data strictly contradict both proposed high-temperature
structures: the linear spin chain \cite{Carpy:75} as well as the symmetric quarter-filled ladder
\cite{Smo:98,vonSchn:98,Mee:98}.

Recently, detailed optical investigations by Damascelli {\em et al.} \cite{Damascelli:00} revealed
a local charge disproportion on each rung of the ladders, which also destroys the inversion
symmetry yielding a finite DM interaction. Hence the charge distribution has to be taken into
account more carefully (cf. Fig. \ref{fig3}c). Following the analysis of Damascelli, the center of
the electronic distribution $x_{\rm{e}}$ on the rungs of the ladder can be determined. Considering
each rung as an independent linear molecule of two V$^{5+}$ ions, the eigenstates of the additional
electron in between are given by $\psi_1 = u \psi_{\rm{l}} + v \psi_{\rm{r}}$, where the electron
stays near the left-hand side and $\psi_2 = u \psi_{\rm{r}} - v \psi_{\rm{l}}$, where the electron
favors the right-hand side. The atomic orbitals $\psi_{\rm{r}}$ and $\psi_{\rm{l}}$ of the
right-hand and left-hand side are occupied according to the following weights:
\begin{equation}
u = \frac{1}{\sqrt
2}\sqrt{1+\frac{\Delta}{E_{\rm{CT}}}},\hspace{0.5cm} v =
\frac{1}{\sqrt
2}\sqrt{1-\frac{\Delta}{E_{\rm{CT}}}}\label{weights}
\end{equation}
The parameter $\Delta$ describes the asymmetry of the onsite energy of the electron located in
$\psi_{\rm{r}}$ with respect to $\psi_{\rm{l}}$. The charge-transfer energy
$E_{\rm{CT}}=\sqrt{\Delta^2+4t^2}$ is determined by the hopping integral $t$ and resembles the
energy splitting between the two eigenstates. We choose the origin of the coordinate system $x=0$
in the left-hand ion, giving the position of the right-hand ion at $x=R$. If the electron favors
for example the left-hand side, the center of the charge distribution $x_{\rm{e}}$ can be obtained
from the eigenstate $\psi_1$ as
\begin{equation}
x_{\rm{e}}= \frac{v^2 R + (R/2)2uv S_{\rm{RL}}}{u^2 + v^2 + 2uv
S_{\rm{RL}}} \approx \frac{R}{2}
(1-\frac{\Delta}{E_{\rm{CT}}})\label{cm}
\end{equation}
where $S_{\rm{RL}}$ is the overlap integral between $\psi_{\rm{r}}$ and $\psi_{\rm{l}}$. The last
expression was obtained assuming $S_{\rm{RL}} \ll 1$, which is justified with respect to the
distance $R\approx 3.6\,{\rm\AA}$ of both ions, and using equation \ref{weights}. Taking the
experimental values $\Delta\approx 0.8$\,eV and $t\approx 0.3$\,eV from \cite{Damascelli:00}, we
calculate $x_{\rm{e}}=0.1 R \approx 0.36\,{\rm\AA}$, yielding a shift of the electronic charge
distribution by $0.36\,{\rm\AA}$ with respect to the center of the left-side V$^{5+}$ ion. If now
neighboring rungs of the ladder locally obey the same charge distribution, we obtain an overall
shift of $\Delta a\approx 0.64\,{\rm\AA}$ of the vanadium spins with respect to the connecting
oxygen ion. This yields an V-O-V bridge angle of about $140^{\circ}$ within the $a$-$b$ plane and
therefore a dominant DM contribution in $c$ direction. The EPR data provide strong experimental
evidence for the appearance of charge disproportions in the quarter-filled spin ladder at high
temperatures.

The peculiarities observed in the cross-over regime (34\,K $<T<60$\,K) can be described in the
framework of the competition of antisymmetric DM exchange and symmetric anisotropic exchange, which
has been discussed by Yamada {\em et al.} \cite{Yam:89} for the one dimensional antiferromagnet
KCuF$_3$ and more extensively for ferromagnetic Cu layers by Soos {\em et al.} \cite{Soos:77}. In
these papers the evolution of the linewidth with temperature is calculated for both interactions.
In both cases the linewidth saturates at high temperatures: The antisymmmetric DM exchange yields a
strong temperature dependence of the linewidth with a negative curvature within the whole
temperature regime. The anisotropic exchange gives rise only to a weak temperature dependence of
the linewidth with a positive curvature at low temperatures changing to a negative one at high
temperatures. Combining both contributions, we qualitatively obtain the temperature dependence of
the linewidth observed in NaV$_2$O$_5$ experimentally. Comparing these data to those of the one
dimensional Heisenberg antiferromagnet KCuF$_3$ in more detail, we recognize an important
difference. Whereas in KCuF$_3$, in agreement with theoretical predictions, the DM interaction
vanishes exactly at $T=0$, for NaV$_2$O$_5$ the extrapolation of the high-temperature data suggests
the DM interaction to disappear already at 100\,K, as documented in the lower inset of figure
\ref{fig1}. This indicates that already for temperatures $T \leq$ 100\,K the charge disproportions
change to a symmetric zig-zag order of V$^{4+}$ and V$^{5+}$ ions in the ladder, where the DM
vectors cancel each other (cf. Fig. \ref{fig3}d). This result is in good agreement with reports of
charge-order fluctuations for 80\,K $\geq T\geq T_{\rm{SP}}$ from Raman spectroscopy \cite{Fis:99}
and the onset of a static long-range zig-zag order below $T_{\rm{SP}}$ for ultrasonic
\cite{Schwenk:99} and dielectric results \cite{Smirnov:99}.

Finally the $\pi/2$-periodic modulation of the angular dependence in the $b$-$c$ plane according to
equation \ref{aniso} can be understood in terms of the anisotropic exchange, only. The DM
interaction produces $\pi$-periodic modulations in any case. The anisotropic-exchange interaction
yields $\pi/2$-periodic modulations only if its secular contributions are enhanced with respect to
the non-secular contributions by spin diffusion, which is characteristic for low dimensional
systems \cite{Soos:77}. As we will report in a more detailed subsequent publication, the fact that
the $\pi/2$-periodic modulation is only observed in the $b$-$c$ plane, whereas it does not occur in
both $a$-$b$ and $a$-$c$ plane, means that all elements of the anisotropic exchange tensor vanish
except $J_{\rm{bc}}\neq 0$. Under these conditions, the prefactors of equation \ref{aniso} resemble
anisotropic $A\sim J_{\rm{bc}}^2$ and antisymmetric $B\sim A_{\rm{DM}}$ exchange. From the lower
inset of figure \ref{fig2} we obtain that for 60\,K $>T>34$\,K the Dzyaloshinsky-Moriya interaction
more and more decreases with decreasing temperature whereas the anisotropic exchange slightly
increases. The anisotropic exchange parameter $J_{\rm{bc}}$ is due to the coupling between electron
spins of adjacent layers in the $c$ direction. This is a strong hint that the ordering below
$T_{\rm{SP}}$ involves vanadium ions of neighboring layers.\\

In conclusion the EPR data confirm the structure of a quarter-filled spin ladder with strong local
charge disproportions at high temperatures $T>100$\,K, which result in a non-vanishing DM
interaction (Fig. \ref{fig3}c). The direction of the DM vector is determined by the V-O-V bridge
angle $\approx 140^{\circ}$ along the chains within the $a$-$b$ plane. The pronounced weakening of
the DM interaction below 100\,K but far above $T_{\rm{SP}} = 34$\,K indicates the onset of zig-zag
charge-order fluctuations, which become long-range and static below $T_{\rm{SP}}$ (Fig.
\ref{fig3}d). The zig-zag structure probably results from inter-layer couplings.\\

This work was supported by the Bundesministerium f\"{u}r Bildung und Forschung (BMBF) under Contract
No. EKM 13N6917/0 and by the Sonderforschungsbereich 484 of the Deutsche Forschungsgemeinschaft.
M.~V.~E.~was partially supported by RFFI Grant - 00-02-17597.

\end{multicols}

\end{document}